\documentclass[11pt,prd,preprintnumbers,nofootinbib,superscriptaddress,notitlepage]{revtex4-1}

\usepackage{graphics}
\usepackage{bm}
\usepackage{cancel}
\usepackage{color}
\usepackage{xcolor}
\usepackage{multirow}
\usepackage{slashed}
\usepackage{array}
\usepackage{amssymb}
\usepackage{graphicx}
\usepackage{mathrsfs}
\usepackage{diagbox}
\usepackage{amsmath}
\usepackage{float}
\usepackage{extarrows}
\usepackage[colorlinks,citecolor=blue,anchorcolor=red,menucolor=red,linkcolor=red,filecolor=red,runcolor=red,urlcolor=red,frenchlinks=red]{hyperref}
\usepackage{subfigure}
\usepackage{ulem}

\makeatletter
\newcommand{\thickhline}{\noalign {\ifnum 0=`}\fi \hrule height 1pt\futurelet \reserved@a \@xhline}
\newcolumntype{"}{@{\hskip\tabcolsep\vrule width 1pt\hskip\tabcolsep}}
\makeatother

\newcommand{\red}{\rm Red}
\newcommand{\rea}{\rm Rea}
\newcommand{\imd}{\rm Imd}
\newcommand{\ima}{\rm Ima}

\newcommand{\ee}{$e^+e^-$}
\newcommand{\tautau}{$\tau^+\tau^-$}
\newcommand{\OO}{$\mathcal{O}$}
\newcommand{\oo}{\mathcal{O}}
\newcommand{\re}{\text{Re}}
\newcommand{\im}{\text{Im}}

 \newcommand{\kk}{\boldsymbol{\rm k}}

 \newcommand{\hp}{\boldsymbol{\hat{\rm p}}}
 \newcommand{\hk}{\boldsymbol{\hat{\rm k}}}

 \newcommand{\sip}{\boldsymbol{\sigma}_+}
 \newcommand{\ssim}{\boldsymbol{\sigma}_-}
 
\begin{document}

\title{Probing the electromagnetic dipole moment of the $\tau$ lepton in the $e^+e^- \to \gamma^*/Z \to \tau^+ \tau^-$ reaction }

\author{Peng-Cheng Lu}
\email{pclu@sdu.edu.cn}
\affiliation{School of Physics, Shandong University, Jinan, Shandong 250100, China}

\author{Zong-Guo Si}
\email{zgsi@sdu.edu.cn}
\affiliation{School of Physics, Shandong University, Jinan, Shandong 250100, China}

\author{Han Zhang}
\email{ han.zhang@mail.sdu.edu.cn}
\affiliation{School of Physics, Shandong University, Jinan, Shandong 250100, China}

\begin{abstract}

High-precision measurements of the lepton's electromagnetic dipole moments provide a powerful probe for testing the Standard Model.
A non-zero value of the  $\tau$ lepton's electric(weak) dipole moment ($d_{\tau}^{\gamma}$, $d_{\tau}^{Z}$) would serve as a smoking gun of the new physics. On the one hand, current and future high energy colliders offer an ideal environment for such measurements. On the other hand, it is essential to investigate the optimal measurement method for extracting not only $d_{\tau}^{\gamma}$ and $d_{\tau}^{Z}$, but also the anomalous magnetic dipole moment ($a_{\tau}$). In this work, we analyze the precision of observables, particularly optimal observables, for determining these physics quantities through the process $e^+e^- \to \gamma^*/Z \to \tau^+ \tau^-$ with $\tau$ leptons undergoing (semi-)leptonic and hadronic decays at the centre-of-mass energies at the Z pole mass and significantly below $m_Z$. By considering the full kinematic information of the decay products, we find that the sensitivities to Im$d_{\tau}$ and Re$d_{\tau}$ can reach $10^{-21}$ $ecm$ at CEPC  $\sqrt s = m_Z$, compared to $10^{-20}$ $ecm$ at Belle-II $\Upsilon(4S)$ resonance and $10^{-17}$ $ecm$ at BEPCII $\psi(2S)$ resonance. For $a_{\tau}$, we find that Im$a_{\tau}$ and Re$a_{\tau}$ with a precision of $10^{-5}$ at Belle-II $\sqrt s=10.58$ GeV can be attained.

\end{abstract}

\maketitle

\section{INTRODUCTION}\label{sec1}

The search for electric dipole moments (EDM) and weak dipole moments (WDM)  of fundamental fermions is an important aspect of experimental investigations  hunting for physics beyond the Standard Model (SM) of particle physics, in particular for Charge-Parity (CP) violation beyond the Kobayashi-Maskawa mechanism (KM) \cite{Kobayashi:1973fv}. In SM, EDMs and WDMs of leptons induced by loop corrections are too small ($|d_{\tau}| \leq 10^{-34}$ $ecm$) \cite{Hoogeveen:1990cb} to be observed by current and near-future experiments. Any observation of a non-zero value for the EDM and/or WDM of a elementary particle would be a smoking gun evidence of non-SM sources of CP violation.  Stringent upper limits on the EDM of the electron and muon have been obtained which are $|d_e|\le 4.1\times 10^{-30}$ $ecm$ \cite{Roussy:2022cmp} and $|d_\mu|\le 1.9\times 10^{-19}$ $ecm$ \cite{Muong-2:2008ebm}. Sizable EDMs of heavy fermions i.e. of the $\tau$ lepton can be generated in some models beyond the SM (BSM) \cite{Bernreuther:1996dr}, which might be detectable in the near future. However, due to short mean lifetime of the $\tau$ lepton, around $2.9\times 10^{-13}$s \cite{ParticleDataGroup:2024cfk}, it decays too fast to be detected by using standard spin precession techniques, but searches using other methods can be carried out at the colliders.
The most precise determination of the EDM of the $\tau$ lepton was obtained at Belle using $e^+ e^- \to \tau^+\tau^-$ events with 833 $fb^{-1}$ data sample \cite{Belle:2021ybo},  which read, for the real and imaginary component respectively:
\begin{align}
&\text{Re}[d_{\tau}] = (-6.2 \pm 6.3) \times 10^{-18} ~ ecm, \nonumber \\
&\text{Im}[d_{\tau}] = (-4.0\pm 3.2) \times 10^{-18} ~ ecm.
\end{align}
For the measurement on WDM of the $\tau$ lepton, the current best search results come from LEP, by measuring the transverse and normal $\tau$ lepton polarization \cite{Lohmann:2005im}:
\begin{align}
&\text{Re}[d^Z_{\tau}] = (-0.6.5 \pm 1.49)  \times 10^{-18} ~ecm, \nonumber \\
&\text{Im}[d^Z_{\tau}] = (0.4 \pm 3.8) \times 10^{-18} ~ecm.
\end{align}

The anomalous magnetic dipole moments (AMDM) of leptons, defined as $a_l = (g-2)/2$, are also used as probes of New Physics (NP). In the SM, $a_{\ell}$ is generated by radiative corrections whose leading order (LO) contribution is given by the so-called Schwinger term $a^{\text{(LO,SM)}}_l = \alpha/(2 \pi)$.
$a_e$ shows one of the best agreements  between the SM prediction and the experimental result with precision up to 10 significant digits \cite{Kinoshita:2014lmy,Parker:2018vye}. The Muon g-2 collaboration published the value of $a_{\mu}$ by combining the previous results given in \cite{Muong-2:2006rrc,Muong-2:2021ojo,Muong-2:2023cdq}, which showed a 5.1$\sigma$ discrepancy between the theory prediction \cite{Aoyama:2020ynm} and these experiments \cite{Cotrozzi:2024bmc}.  Based on lattice QCD calculation and a new $\tau$-data-driven analysis, these discrepancies decrease to less than 2$\sigma$ \cite{Toth:2022lsa,Masjuan:2024ccq}. Very recently, this tension was removed in ref.\cite{Aliberti:2025beg}.
Since many BSM theories predict lepton's AMDM corrections scaling as $\delta a_l \propto \frac{m_l^2}{\Lambda^2}$ \cite{Martin:2001st},
the $\tau$ lepton exhibits enhanced sensitivity to the BSM effects compared to lighter leptons. 
Currently, the most stringent constraint for $a_{\tau}$ is obtained at the Large Hadron Collider (LHC) by the CMS collaboration using proton-proton collisions \cite{CMS:2024qjo}. The resulting constraint on $a_{\tau}$ reads 
\begin{align}
-0.0042 < a_{\tau} < 0.0062,~~ 95\% ~\text{CL}.
\end{align}
On the other hand, the SM prediction is $a^{\text{SM}}_{\tau} = 117717.1 \pm 3.9 \times 10^{-8}$ \cite{Keshavarzi:2019abf}.

The high-energy, high-luminosity colliders - notably Belle-II at SuperKEKB \cite{Belle-II:2010dht,Belle-II:2018jsg}, Beijing Electron-Positron Collider II (BEPCII) \cite{BESIII:2024lks,BESIII:2022mxl}, and the future Circular Electron-Positron Collider (CEPC)\cite{CEPCStudyGroup:2018ghi,CEPCPhysicsStudyGroup:2022uwl} - will enable precision measurements of the $\tau$ lepton's electromagnetic dipole moments through a huge amount of $\tau$-pair production.
The reaction $e^+ e^- \to \tau^+\tau^-$ is particularly suitable for probing these quantities when colliders run at the centre-of-mass (c.m.) energies of Z pole or significantly below $m_Z$. Assuming the scale of NP lies above the energy transfer of the process, the effective field theory allows the $\gamma \tau \tau$ and $Z\tau \tau$ vertex modifications, measured via this reaction, to be interpreted as the constrains on the $d_{\tau}$ and $a_{\tau}$. 

A possible way to improve the measurement precision of small physical parameters, i.e. $d_{\tau}$ and $a_{\tau}$, involves constructing proper experimental observables with enhanced sensitivity. While straightforward observables utilize the kinematic variables of the detected final states, optimal observable minimize the statistical uncertainties in the measurement through maximum likelihood estimation \cite{Atwood:1991ka, Diehl:1993br}. 
 Such observables have been extensively used in previous studies, including searches for the CP properties in Higgs boson interactions at LHC \cite{ATLAS:2020evk,ATLAS:2023mqy,CMS:2021nnc} and studies of anomalous charged triple gauge couplings \cite{Jahedi:2024wnw,OPAL:2003xqq}.
In the context of probing the E(W)DM and AMDM of $\tau$ lepton, ref.\cite{Bernreuther:2021uqm} investigated both real and imaginary components of $\tau$ EDM in $\tau$-pair production via \ee collisions at Belle-II; Ref.\cite{Chen:2018cxt} explored the EDM and AMDM of $\tau$ lepton at same energy region while neglecting the imaginary components; Ref.\cite{He:2025ewk} studied $d_{\tau}^{\gamma}$ at low-energy \ee colliders using one-prong $\tau$ lepton decays channels; Ref.\cite{Bhide:2024nje} probed optimal measurements of electromagnetic dipole moments of the $\tau$ lepton in PbPb collisions at LHC. Additionally, ref.\cite{Banerjee:2023qjc} utilized angle distributions to analyze the impact of the real component of $\tau$ EDM and AMDM at a c.m. energy of 10.58 GeV. 

In this study, we investigate the E(M)DM and AMDM of the $\tau$ lepton through $e^+ e^- \to \gamma^*/Z \to \tau^+\tau^-$ process, followed by the subsequent  (semi-)leptonic and hadronic decays of the $\tau$ leptons, and estimate the sensitivity for these dipole moments at both the Z pole energy and lower energies in $e^+e^-$ collisions. This paper is structured as follows. In section \ref{sec2} we derive the production and decay spin density matrix for $e^+ e^- \to \tau^+\tau^-$ incorporating the $\tau$ lepton's (semi-)leptonic and hadronic decay channels.  Recalling the form factor decomposition of the $\tau^+\tau^-V$ vertex, (V = $\gamma^*$, Z) and its relevant symmetry properties, we then turn to the observables in section  \ref{sec3}, constructing simple and optimal observables to probe the electromagnetic dipole moment and analyzing the expectation values and covariances of these observables.  Section  \ref{sec4} presents our numerical results, including the estimates of the sensitivity for the measurement of real and imaginary components of E(M)DM and AMDM using observables defined in section \ref{sec3}. Comparisons are made across collider energy regimes.  We summarize in Section  \ref{sec5}. The analytical expressions for the production spin density matrix of this process are documented in the Appendix.

\section{The reaction}\label{sec2}

We consider \tautau production in \ee collisions via a virtual $\gamma$ and Z boson,
decaying into  final states A and $\bar B$, respectively,
\begin{align}
\label{reaction1}
e^-(p_1) + e^+(p_2) \to \tau^-(k_1,\alpha) + \tau^+(k_2, \beta) \to A + \bar B.
\end{align}
The four-momenta and the corresponding three-momenta of electron and $\tau$ leptons are denoted in the \ee ~c.m. frame by $p_{1,2}=(p^0_{1,2},\bold{p_{1,2}})$, $k_{1,2}=(k^0_{1,2},\bold{k_{1,2}})$.
We consider unpolarized electrons and positrons and neglect their masses, the labels
$\alpha, \beta$ denote the spin indices of the $\tau$ leptons. 
In this process, the general form factors for $\tau \tau \gamma$ and $\tau \tau Z$ vertex functions  can be  written in the form
\begin{align}\label{eq:coupling}
\Gamma^{\mu}(q^2) &= -i e Q_{\tau} (\gamma^{\mu}  + \frac{\sigma_{\mu \nu}q^{\nu}}{2m}[i F_1(q^2) + F_2(q^2) \gamma_5] ) \\ \nonumber
&+ \frac{g_Z}{2} (\gamma^{\mu}(V_{\tau} - \gamma_5 A_{\tau})  + \frac{\sigma_{\mu \nu}q^{\nu}}{2m}[i G_1(q^2) + G_2(q^2) \gamma_5] ),
\end{align}
where $m$ is the mass  of the $\tau$ lepton and $e Q_{\tau}$ is the  $\tau^-$ charge, $g_Z $ is weak coupling constant,  $\sigma_{\mu \nu} = i[\gamma_{\mu},\gamma_{\nu}]/2$ and $q = k_{1} + k_{2}$. The couplings with the form factors $F_1(q^2)$ and $G_1(q^2)$ are CP conserving, while those with $F_2(q^2)$ and $G_2(q^2)$ are CP violating.
 In the limit of $q^2 \to 0$, the form factor $F_1(q^2)$ is directly related to electric anomalous magnetic dipole moment
\begin{align}
F_1(0) = a^{\gamma}_{\tau}=(g-2)/2,
\end{align}
while the Pauli form factor $F_2(q^2)$ describes the EDM of $\tau^{\pm}$
\begin{align}
\frac{e Q_{\tau}}{2m}F_2(0) = d^{\gamma}_{\tau}.
\end{align}
$G_1(0)$ and $G_2(0)$ are related to the weak anomalous magnetic moment and weak dipole moments of $\tau^{\pm}$, which are denoted respectively  by $a_{\tau}^Z$ and  $d_{\tau}^Z$ in the following.  In our analysis, we  shall consider the $a_{\tau}^V$ and $d_{\tau}^V$ (V = $\gamma^*$, Z) to be complex, with their imaginary parts being connected with the absorptive part of the vertex.

At tree level in the SM, $a_{\tau}^V = 0$.  Anomalous magnetic moment can have contributions both from NP and electroweak radiative corrections in the SM, thus $a_{\tau}^V = a_{\tau}^{\text{SM}} + a_{\tau}^{\text{NP}}$. 
As the electric dipole form factor in the SM is highly suppressed, one can assume that this form factor comes exclusively from NP contributions, that is $d^V_{\tau} = d^{\text{NP}}_{\tau} $. 

For our calculations, we used the coordinate system where the electron beam is chosen as $z$-axis and one orthogonal direction as $x$-axis in the c.m. frame of the \ee and the polar coordinate for the decay products of the $\tau$ leptons in the respective $\tau$ rest frames that is defined by a rotation-free Lorentz boost from the c.m. \ee. In the $\tau^-$($\tau^+$) rest frame, the spin of the $\tau^-$($\tau^+$) is described by a unit vector $\hat{s}_1$($\hat{s}_2$).  With the aid of Lorentz boost transformation, we have the components value of Pauli-Lubanski-4-vector $s^{\mu}_1$ and $s^{\mu}_2$ in the c.m. \ee frame 
\begin{align}
s^{\mu}_1 &= (\frac{\bold{k_1} \cdot  \hat{s}_1}{m},  \hat{s}_1 + \frac{\bold{k_1} (\bold{k_1} \cdot  \hat{s}_1)}{m (m+k_0)}  ),\\ \nonumber
s^{\mu}_2 &= (- \frac{\bold{k_1} \cdot  \hat{s}_2}{m},  \hat{s}_2 + \frac{\bold{k_1} (\bold{k_1} \cdot  \hat{s}_2)}{m (m+k_0)}  ),
\end{align}
with $s^2_1=s^2_2=-1$ and $k_1 \cdot s_1 = k_2 \cdot s_2 = 0$, where $k_0$ is the energy of the $\tau^-$ in the \ee c.m. frame.

For on-shell $\tau$ pair production and decay in the reaction (\ref{reaction1}), the total cross section with eq.(\ref{eq:coupling}) for $\tau \tau V$ couplings is given by the product of the production density matrix $ \chi_{\alpha \alpha' \beta \beta'}$ for \ee $\to$ \tautau and the density matrices $\mathcal{D}^a_{\beta' \beta}$ and $\mathcal{D}^{\bar{b}}_{\alpha' \alpha}$ that describe the decays of polarized $\tau^- \to A \to a(q_-) +X$ and $\tau^+ \to \bar{B} \to \bar{b}(q_+) + X'$, respectively. The narrow width approximation of the intermediate $\tau$ leptons is used. Thus the differential cross section can be written as 
\begin{align}
d\sigma_{a\bar{b}} =\frac{1}{2s} \chi_{\alpha \alpha' \beta \beta'} \mathcal{D}^a_{\beta' \beta} \mathcal{D}^{\bar{b}}_{\alpha' \alpha}  d\phi,
\end{align}
where $s=(p_1 + p_2)^2$ and $d\phi$ is the usual phase space measure. 
A useful parametrization of $\chi_{\alpha \alpha' \beta \beta'}$ is the Fano-Bloch decomposition \cite{Fano:1983zz}
\begin{align}
\chi_{\alpha \alpha' \beta \beta'} = A \delta_{\alpha \alpha'} \delta_{\beta \beta'} + B_{1i} \sigma^i_{\alpha \alpha'} \delta_{\beta \beta'} + B_{2 i} \delta_{\alpha \alpha'}\sigma^i_{\beta \beta'} + C_{ij} \sigma^i_{\alpha \alpha'}\sigma^j_{\beta \beta'},
\end{align}
where $\sigma^i$ are the Pauli matrices with $i,j=1,2,3$. The function $A$ is related to the spin-averaged production cross section. The function $B_{1i}$ and $B_{2i}$ are associated with the net polarization of the $\tau^-$ and $\tau^+$ and the function $C_{ij}$ encode the spin correlation.

We expand the production density matrix of the $\tau$ lepton to first order in the dipole and anomalous magnetic moment form factors, neglecting the spin indices:
\begin{align}
\label{eq:densityM}
\chi_{prod} = \chi_{0} + \text{Re}[d_{\tau}]~ \chi_{Red} + \text{Im}[d_{\tau}]~ \chi_{Imd} + \text{Re}[a^{NP}_{\tau}] ~\chi_{Rea} +  \text{Im}[a^{NP}_{\tau}] ~\chi_{Ima} + \mathcal{O}(a_{\tau}^2,d_{\tau}^2),
\end{align}
where $\chi_{0}$ is the SM term where $d^{NP}_{\tau} = 0$ and $a^{NP}_{\tau} = 0$. $\chi_{Red}$ and  $\chi_{Imd}$ are the interference terms between SM and NP-induced E(W)DM for the real and imaginary components of $d_{\tau}$, while $\chi_{Rea}$ and  $\chi_{Ima}$ are the interference terms between SM and NP-induced AMDM for the real and imaginary components of $a_{\tau}$. We neglect the higher order term of $d^V_{\tau}$ and $a^V_{\tau}$ since they are small.

The specific expressions of $\chi_i$($i= Red, Imd, Rea, Ima$) as well as tree level SM density matrix $\chi_{0,tree}$  are presented in the appendix. The  expressions given do not include the interference terms between the Z and $\gamma$ because we only consider the $e^+e^-$ c.m. energy at the lower energy and Z pole energy.  The tree level $\chi_{0,tree}$  is CP even and Naive-CPT even. Once the electroweak correction is considered, this term may become Naive-CPT odd but CP even.  
The density matrix $\chi_{Red}$ and $\chi_{Imd}$ are CP odd, with $\chi_{Red}$ being T even and $\chi_{Imd}$ being T odd. For the density matrix associated with AMDM, $\chi_{Rea}$ is CP even and T even, while $\chi_{Ima}$ is CP even and T odd.
T even or odd refers to the behavior under naive time reversal transformation, which reverses particle momenta and spins but does not interchange initial and final states.

For the decay channels, we consider $\tau$ pair events in two cases:
\begin{itemize}
\item One-prong decays: Only one charged particle from each of $A$ and $\bar B$ is measured. Including
 $\tau^- \to \ell^-(q_{-}) \nu \nu$ ($\ell = e, \mu$), $\pi^-(q_{-}) \nu$, $\rho^-(q_{-}) \nu$, along with the charge-conjugate $\tau^+$ decay.
\item Multi-prong decays: More than one particle from each of $A$ and $\bar B$ are measured. Including
  $\tau^- \to \pi^-(q_{1}) \pi^0(q_2) \nu$, $\pi^-(q_{1}) \pi^-(q_2) \pi^+(q_3) \nu$, along with the charge-conjugate $\tau^+$ decay.
\end{itemize}

After integrating the energy of the charged particle $a^{\pm}$ that acts as $\tau^{\pm}$ spin analyzer, the decay density matrix in the $\tau$ rest frame for one charged prong case can be written as 
\begin{align}
\frac{d^3 q_{\pm}}{(2 \pi)^3 2 q_{\pm}^0} \mathcal{D}_{\beta^{\prime} \beta}^{a^{\pm}}\left[\tau^{\pm} \to a^{\pm}\left(q_{\pm}\right)+X\right]  =\frac{d\Omega_{a^{\pm}}}{4\pi} [1 + \alpha_a \hat{q}_{\pm}\cdot \sigma],
\end{align}
where $d\Omega_{a^{\pm}}$ is the solid angle of $a^{\pm}$. $\alpha_a$ is the spin-analying power of $\tau$ lepton. In the SM, it is well known that the $\pi$ and $\rho$ have high spin analyzing power, with $\alpha_{\pi} = 1$ and $\alpha_{\rho} = 0.45$. The decay density matrices for polarized $\tau^{\pm}$ in several major decay channels have been calculated in \cite{Bernreuther:2021elu}.

\section{ Observables}\label{sec3}

In this section, we discuss the simple observables and the so-called optimal observables, along with their sensitivities for determining the values of $d_{\tau}$ and $a_{\tau}^{NP}$. We decompose the differential cross section of the reaction(\ref{reaction1}) according to $g_j$, neglecting terms quadratic in $g_j$:
\begin{align}
\label{eq:dcs}
d\sigma_{a\bar b} = (S_0 + S_{1,Red_{\tau}} \re[d_{\tau}] + S_{1,Imd_{\tau}} \im[d_{\tau}] +S_{1,Rea_{\tau}} \re[a_{\tau}] +S_{1,Ima_{\tau}} \im[a_{\tau}]) d\phi,
\end{align}
here $g_j$($j=1,2,3,4$) represents the quantities $\text{Re}[d_{\tau}], \text{Im}[d_{\tau}], \text{Re}[a^{NP}_{\tau}], \text{and} ~\text{Im}[a^{NP}_{\tau}]$ to be determined. 
To determine these quantities, one needs to choose four observables $\mathcal{O}_i$ and measure their expectation values in the same  experimental run:
\begin{align}
E[\mathcal{O}_i]=  E_0[\mathcal{O}_i] + \sum_j c_{ij} g_j,
\end{align}
$E_0[\mathcal{O}_i] $ denotes the expectation value without contributions from $g_j$ and  $c_{ij}$ is the coefficient matrix that specifies the relationship between $E[\oo_i]$ and the $g_j$. 
The goal is to measure $g_j$ as accurately as possible. Consider a set of unbiased estimators, denoted as $\gamma_i$, defined as
 $\gamma_i = \sum_j c^{-1}_{ij}(E[\mathcal{O}_i] -E_0[\mathcal{O}_i] )$. The expectation values for these estimators are $E[\gamma_i] = g_i$.  Expanding in $g_j$ gives 
\begin{align}
\label{eq:16}
E[\mathcal{O}_i] = \frac{\int d\phi S_{0} \oo_i}{\int d\phi S_{0}} + \sum_j \left( \frac{\int d\phi S_{1,j} \oo_i}{\int d\phi S_0} -  \frac{\int d\phi S_0 \oo_i \int d\phi S_{1,j}}{(\int d\phi S_0)^2}\right) g_j.
\end{align}
To further assess the sensitivity of observables to those small couplings, one must compare with the errors in the measurement of expectation of observables. If $N$ events are recorded, these error will be given by $\sqrt {V(\oo)_{ij}/N}$, with 
\begin{align}
\label{eq:covariance}
V(\oo)_{ij} = \frac{d\phi S_0 \oo_i \oo_j}{\int d\phi S_0} - \frac{\int d\phi S_0 \oo_i \int d\phi S_0 \oo_j}{(\int d\phi S_0)^2}
\end{align}
being the covariance matrix of the quantities \OO ~for $g_j=0$. This matrix is symmetric and positive semi-definite and will be a diagonal matrix in the case of all \OO ~being un-correlated observables. Thus, the `one-sigma' standard deviation or the so-called background-to-signal ratio $\delta g_i$, with which the presence of non-zero $g_i$ can be ascertained reads
\begin{align}
\label{eq:1sd}
\delta g_i = \frac{1}{|c_{ij}|} \sqrt{\frac{V(\oo)_{ij}}{N}}.
\end{align}

We now discuss the four optimal observables which reads
\begin{align}
\label{eq:Ore}
\oo_i = \frac{S_{1,i}}{S_0}, (i = Red_{\tau}, Imd_{\tau}, Rea_{\tau}, Ima_{\tau})
\end{align}
that can optimize the signal-to-background ratio in the measurements \cite{Atwood:1991ka, Diehl:1993br}. 
$S_{1,i}$ and $S_0$ are the differential cross section defined by eq.(\ref{eq:dcs}). Clearly, optimal observables inherit whatever discrete asymmetries $S_{1,i}$ has with respect to $S_0$. In this work, it is suitable to define observables especially CP odd observables in terms of process (\ref{reaction1}) and its charge conjugated decays. The expectation value of the observables of interest are defined as the following average \begin{align}
\label{eq:generalO}
E^{ab}[\mathcal{O}]= \frac{1}{2} (\frac{\int d\sigma_{a\bar{b}}\mathcal{O}}{\int d\sigma_{a\bar{b}}} + \frac{\int d\sigma_{b\bar{a}}\mathcal{O}}{\int d\sigma_{b\bar{a}}} ),
\end{align}
where $d\sigma_{b\bar a}$ is the differential cross section for the charge-conjugate process of reaction(\ref{reaction1}).
These optimal observables share the same transformation properties under CP and T as the couplings. Specifically, observables sensitive to the $\gamma \tau \tau$ couplings will also be sensitive to the corresponding $Z \tau \tau$ couplings.
For example, the expectation value of CP odd observables can only project CP odd term of the differential cross section in eq.(\ref{eq:dcs}). Consequently, for CP even SM contribution, any non-zero expectation value for a CP odd observable serves as a clear signal of CP violation in the reaction. The reverse is also true: CP even term of differential cross section do not contribute to the expectation value of CP odd observable. The discussion here assumes the phase space cuts in experimental measurements are CP blind.

In addition to optimal observables, we also study simple observables constructed solely from the momenta of the decay products of \tautau. For the case of the imaginary and real E(W)DM couplings, we choose simple observables of the form
\begin{align}
\label{T33}
&T^{ij} = (\bold{q}_+ - \bold{q}_-)^i ( \bold{q}_+ \times  \bold{q}_-)^j + (i \leftrightarrow j), ~~ \text{odd under CP; odd under T}, \\ \nonumber
&\hat{O}^{ij}= (\hat{ \bold{q}}_+ + \hat{ \bold{q}}_-)^i (\hat{ \bold{q}}_+ - \hat{ \bold{q}}_-)^j + (i \leftrightarrow j), ~~ \text{odd under CP; even under T},
\end{align}
which have the same symmetry as $S_{1,Red_{\tau}}$ and $S_{1,Imd_{\tau}}$. Here, $ \bold{q}_{\pm}$ are the three-momenta of the final states defined in the $e^+e^-$ frame, $\hat{ \bold{q}}_{\pm} =  \bold{q}_{\pm}/| \bold{q}_{\pm}|$ and $i,j$ are the Cartesian vector indices.
For the case of real and imaginary AMDM couplings, we consider simple observables of the form
\begin{align}
\label{Ta33}
T_a^{ij} = ( \bold{q}_+ +  \bold{q}_-)^i ( \bold{q}_+ \times  \bold{q}_-)^j + (i \leftrightarrow j), ~~\text{even under CP; odd under T},
\end{align}
and 
\begin{align}
\label{Oa33}
\hat{O}^{ij}_a = \hat{ \bold{q}}^i_+ \cdot \hat{ \bold{q}}^j_- + (i \leftrightarrow j), ~~\text{even under CP; even under T},
\end{align}
which have the same symmetry as $S_{1,Ima_{\tau}}$ and $S_{1,Rea_{\tau}}$.

\section{Numerical Results}\label{sec4}
We consider \tautau production and their decay in \ee collisions and compute the expectation values of the simple and optimal observables defined in the previous section. We also discuss the one-sigma sensitivities to the E(W)DM and AMDM form factors $\text{Red}_{\tau}, \text{Imd}_{\tau}, \text{Rea}_{\tau} ~\text{and Ima}_{\tau}$.
These expectation values are computed using the production density matrix in eq.(\ref{eq:densityM}) and $\tau$ leptons decay density matrix. No phase space cuts are applied. 

In the case $a=b$, as discussed above, for the CP odd observables, only the term corresponding to the CP odd differential cross section  contributes to their expectation value. This is a direct consequence of the CP symmetry properties, which imply 
\begin{align}
&E_0^{a\bar a}[\oo_{Red_{\tau}}] = 0,  ~~ E_0^{a\bar a}[\oo_{Imd_{\tau}}] = 0,\\
&E_0^{a\bar a}[T^{ij}] = 0,  ~~ E_0^{a\bar a}[\hat{O}^{ij}] = 0.
\end{align}
The cross-term drops out by further applying T symmetry properties, which imply 
\begin{align}
\label{eq:crossE0}
&E_0^{a\bar a}[\oo_{Red_{\tau}} \oo_{Imd_{\tau}}] = 0,\\ \nonumber
&E_0^{a\bar a}[T^{ij} \hat{O}^{ij}] = 0.
\end{align}
Thus, the covariance matrix and coefficient matrix are both diagonal. For cases where $a \neq b$, applying the T transformation results in the following SM expectation values:
\begin{align}
&E_0^{a\bar b}[\oo_{Red_{\tau}}] = 0,  ~~ ~~ E_0^{a\bar b}[T^{ij}] = 0,\\ \nonumber
&E_0^{a\bar b}[\oo_{Red_{\tau}} \oo_{Imd_{\tau}}] = 0,  ~~ E_0^{a\bar b}[T^{ij} \hat{O}^{ij}] = 0,
\end{align}
while $E_0^{a\bar b}[\hat{O}^{ij}] \ne 0$, $E_0^{a\bar b}[\oo_{Imd_{\tau}}] \ne 0$. This behavior arises because the final states are no longer CP symmetric in such cases. Consequently, the average observables defined in  eq.(\ref{eq:generalO}) are used. With the CP property of $\hat{O}^{ij}$ and $\oo_{Imd_{\tau}}$, we obtain
\begin{align}
E_0^{ab}[\hat{O}^{ij}]  = 0,~~ E_0^{ab}[\oo_{Imd_{\tau}}] = 0.
\end{align}
Any non-zero value of $E^{ab}[\oo]$ may be regarded as a potential signal of the NP.

For the process where $\tau$ decays into three $\pi$, the $E_0^{ab}[\oo_{Red_{\tau}} \oo_{Imd_{\tau}}] \neq 0$ is induced by absorptive parts originating from the finite widths of the intermediate resonances. However, this term is significantly smaller  compared to our numerical uncertainties and can thus be safely neglected. Again $E_0^{ab}[\oo_{Red_{\tau}}] = 0$ and  $E_0^{ab}[\oo_{Imd_{\tau}}] = 0$ hold in this 3$\pi$ process.

Now, we consider the expectation values of CP even observables.
 Since the SM expectation value of $\oo_{Ima_{\tau}}$ and $\oo_{Rea_{\tau}}$ are not equal to zero, we utilize the general relationship between expectation values of observables and the values of $a_{\tau}^{NP}$ as defined in eq.(\ref{eq:16}).

Based on the above discussion, the expectation values of the simple observables eq.(\ref{T33}) for E(W)DM and eq.(\ref{Ta33}), eq.(\ref{Oa33}) for AMDM are of the form
\begin{align}
\label{eq:cab}
\left(
\begin{array}{c}
E^{ab}[T^{ij}]\\
E^{ab}[\hat{O}^{ij}]
\end{array}
\right)
=
\left(
\begin{array}{cc}
 c_{ab}  & 0\\
0 & \bar{c}_{ab}
 \end{array}
 \right)
 \left(
 \begin{array}{c}
 \text{Re}[\hat{d}^V_{\tau}]~ s^{ij}\\
\text{ Im}[\hat{d}^V_{\tau}]~ s^{ij}
 \end{array}
 \right),
\end{align}
with dimensionless E(W)DM form factors defined by
\begin{align}
 \text{Re}[\hat{d}^V_{\tau}]= \frac{\sqrt {s}}{e} \text{Re}[d^V_{\tau}] , ~~ \text{ Im}[\hat{d}^V_{\tau}] = \frac{\sqrt {s}}{e} \text{Im}[d_{\tau}] ,
\end{align}
and
\begin{align}
\label{eq:Ta}
\left(
\begin{array}{c}
E^{a b}[T^{ij}_a]\\
E^{a b}[\hat{O}^{ij}_a]
\end{array}
\right)
=
\left(\begin{array}{cc}
k^{11}_{a\bar b}  & k^{12}_{a\bar b}\\
\bar{k}^{21}_{a\bar b} & \bar{k}^{22}_{a\bar b}
 \end{array} \right)
 \left( \begin{array}{c}
 \text{Im}[a^{NP}_{\tau}]~ s^{ij}\\
 \text{Re}[a^{NP}_{\tau}] ~ s^{ij}
 \end{array}\right)
 +
 \left(\begin{array}{c}
 E_0^{ab}[T_a^{ij}]\\
  E_0^{a b}[\hat{O}_a^{ij}]\\
  \end{array}\right)
\end{align}
where $s_{ij}$ is the tensor polarization of the intermediate $\gamma^*/Z$ state, which can be written as $\text{diag}(-\frac{1}{6}, -\frac{1}{6}, \frac{1}{3})$ if the beam direction is identified with the z-axis. 
 When the coefficient matrix is not diagonal in eq.(\ref{eq:Ta}), using the `whitening transformation' to de-correlate the observables, one can find a linearly-approximated optimal basis out of the original chosen ones.
The resulting ideal one standard deviation (1 s.d.) sensitivity of the E(W)DM couplings are given by
\begin{align}
\delta \red_{\tau} = \frac{e}{\sqrt{s}} \frac{1}{\sqrt{N_{ab}}} \frac{3 [E^{ab}_0[T^2_{33}]]^{1/2}}{|c_{ab}|},~~~
\delta \imd_{\tau} = \frac{e}{\sqrt{s}} \frac{1}{\sqrt{N_{ab}}} \frac{3 [E^{ab}_0[\hat{O}^2_{33}]]^{1/2}}{|\bar{c}_{ab}|},
\end{align}
where $N_{ab}=2N_{\tau\tau}Br(\tau\to a)Br(\tau\to b)$ with $N_{\tau\tau}$ being the number of $\tau$ pair events. After the diagonalization operation of the covariance matrix, the 1 s.d. sensitivity of the AMDM couplings can be written as
\begin{align}
\delta \rea_{\tau} = \frac{1}{\sqrt{N_{ab}}} \frac{3 [V^{'}(\hat{O}_{a,33}^{ab})]^{1/2}}{|\kappa_{ab}|},~~~
\delta \ima_{\tau} =  \frac{1}{\sqrt{N_{ab}}} \frac{3 [V^{'}(T_{a,33}^{ab})]^{1/2}}{|\bar{\kappa}_{ab}|},
\end{align}
where $\kappa_{ab}$, $\bar{\kappa}_{ab}$ are components of the new diagonal coefficient matrix and $V^{'}(\hat{O}_{a,33}^{ab})$, $V^{'}(T_{a,33}^{ab})$ are components of the new diagonal covariance matrix.

The expectation values of the optimal observables defined in eq.(\ref{eq:Ore}) are of the form
\begin{align}
\left(
\begin{array}{c}
E^{ab}[\oo_{Red_{\tau}}]\\
E^{ab}[\oo_{Imd_{\tau}}]
\end{array}
\right)
=
\left(
\begin{array}{cc}
 \omega_{ab}  & 0\\
0 & \bar{ \omega}_{ab}
 \end{array}
 \right)
 \left(
 \begin{array}{c}
\text{Re}[\hat{d}^V_{\tau}] \\
\text{ Im}[\hat{d}^V_{\tau}]
 \end{array}
 \right),
\end{align}
and
\begin{align}
\label{eq:Oa}
\left(
\begin{array}{c}
E^{ab}[\oo_{Rea_{\tau}}]\\
E^{ab}[\oo_{Ima_{\tau}}]
\end{array}
\right)
=
\left(\begin{array}{cc}
\xi^{11}_{a\bar b}  & \xi^{12}_{a\bar b}\\
\bar{\xi}^{21}_{a\bar b} & \bar{\xi}^{22}_{a\bar b}
 \end{array} \right)
 \left( \begin{array}{c}
 \text{Im}[a^{NP}_{\tau}] \\
 \text{Re}[a^{NP}_{\tau}] 
 \end{array}\right)
 +
 \left(\begin{array}{c}
E_0^{ab}[\oo_{Rea_{\tau}}]\\
E_0^{ab}[\oo_{Ima_{\tau}}]\\
  \end{array}\right).
\end{align}
The definition of the variables in the above expressions are similar to simple observables.

For the numerical results to be presented below, the following input parameters are adopted:
\begin{align}
m = 1.777 \text{GeV},  m_Z = 91.188 \text{GeV}, m_W = 80.369 \text{GeV}, m_{\pi} = 0.140 \text{GeV},\\ \nonumber
m_{\rho}= 0.775 \text{GeV}, \Gamma_Z =2.496 \text{GeV}, G_{\mu}=1.166379\times 10^{-5} \text{GeV}^{-2}.
\end{align}
The decay channels that we used in this study, along with their branching ratios and spin analyzing powers are shown in Table {\ref{Table:decay-channel}}.
\begin{table}[h]
\caption{Decay channels of $\tau^-$ with the corresponding branching ratios and spin analyzing power.}\label{Table:decay-channel}
\begin{center}
\begin{tabular}{|c |c |c |}
\hline 
 Decay channel		&Spin analyzing power($\alpha_a$)    	& Branching ratio 	 \\ 
\hline
$e^-/\mu^-(q_-) \nu \nu$ 	 	&$-0.33$				&$17.4\%$	 	\\
\hline
$\rho^-(q_-)  \nu$			&$0.45$				&$25.5\%$	 	\\
\hline
$\pi^-(q_-)  \nu$			&$1.00$				&$10.8\%$	 	\\
\hline
$\pi^- (q_1) \pi^0(q_2)  \nu$			&-				&$25.5\%$	 	\\
\hline
$\pi^- (q_1) \pi^-(q_2) \pi^+(q_3) \nu$			&-			&$9.31\%$	 	\\
\hline
\end{tabular}
\end{center}
\end{table}

\subsection{E(W)DM Results}

The EDM  of the $\tau$ lepton can be effectively extracted from $e^+e^- \to \tau^+\tau^-$ at energies significantly lower than the Z boson mass, particularly near the $\tau^+\tau^-$ threshold.
We consider $\tau$ pair production and decay at BEPCII that has been operating since 2009 whose c.m. energies ranging from $\sqrt s=2.0 - 4.94$ GeV with an integrated luminosity of $35 fb^{-1}$ \cite{BESIII:2022mxl}. The largest dataset of the \tautau events are from the $\psi(2S)$ resonance. In 2020, BEPCII plans to make two upgrades. One is increasing the maximum c.m. energy to 5.6 GeV. The other one is increasing the peak luminosity by a factor of 3 for c.m. energies from 4.0 to 5.6 GeV. Besides, we also consider c.m. energy from $\tau$-pair threshold up to 15 GeV. In this energy regime, Belle-II experiment will accumulate about $4.5 \times 10^{10}$ $\tau$ pair at $\Upsilon(4S)$ resonance at $\sqrt s=10.58$ GeV.

WDM of $\tau$ lepton can be best probed experimentally from the $\tau$ pair production at the Z resonance.
The huge amount of produced $\tau^+\tau^-$ pairs at the Z resonance expected at CEPC, together with the improved $\tau$- reconstruction efficiency will allow, among other investigations, the search for $d_{\tau}^Z$ with a precision expected to be siginificantly higher than existing bounds.

In summary, we consider the following four scenarios in this paper:
\begin{itemize}
\item \text{Scenario I }: $\sqrt s$ = 3.686 \text{GeV},~~\text{the number of $\tau^+\tau^-$ }: $N_{\psi(2S)\to \tau^+\tau^-} = 3.5\times 10^6$,\nonumber 
\item \text{Scenario II }: $\sqrt s$ = 5.6 \text{GeV, ~~Integrated luminosity}: $L=20 fb^{-1}$,
\item \text{Scenario III }: $\sqrt s$ = 10.58 \text{GeV}, ~~\text{the number of $\tau^+\tau^-$}: $N_{\Upsilon(4S)\to \tau^+\tau^-} = 4.5 \times 10^{10}$,
\item \text{Scenario IV }: \text{Z-pole energy},~~\text{the number of $\tau^+\tau^-$ }: $N_{Z \to \tau^+\tau^-} = 1.2\times 10^{11}$.
\end{itemize}

\begin{figure}[!h]
\centering
\includegraphics[scale=0.45]{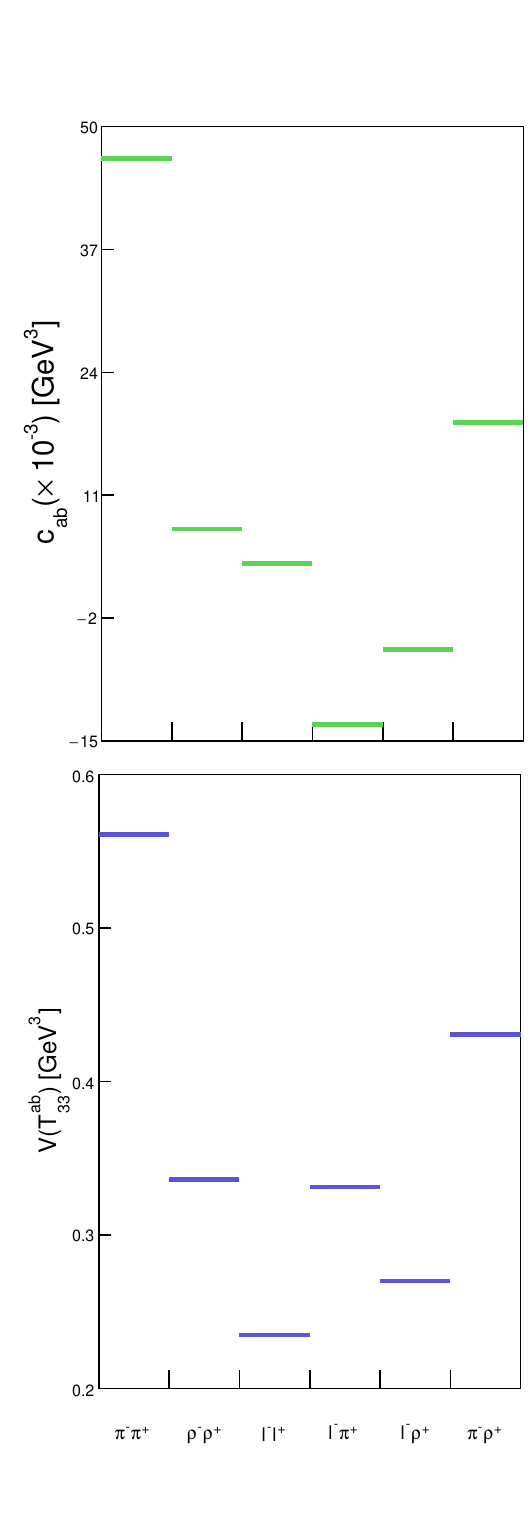}
\hspace{0.4in}
\includegraphics[scale=0.45]{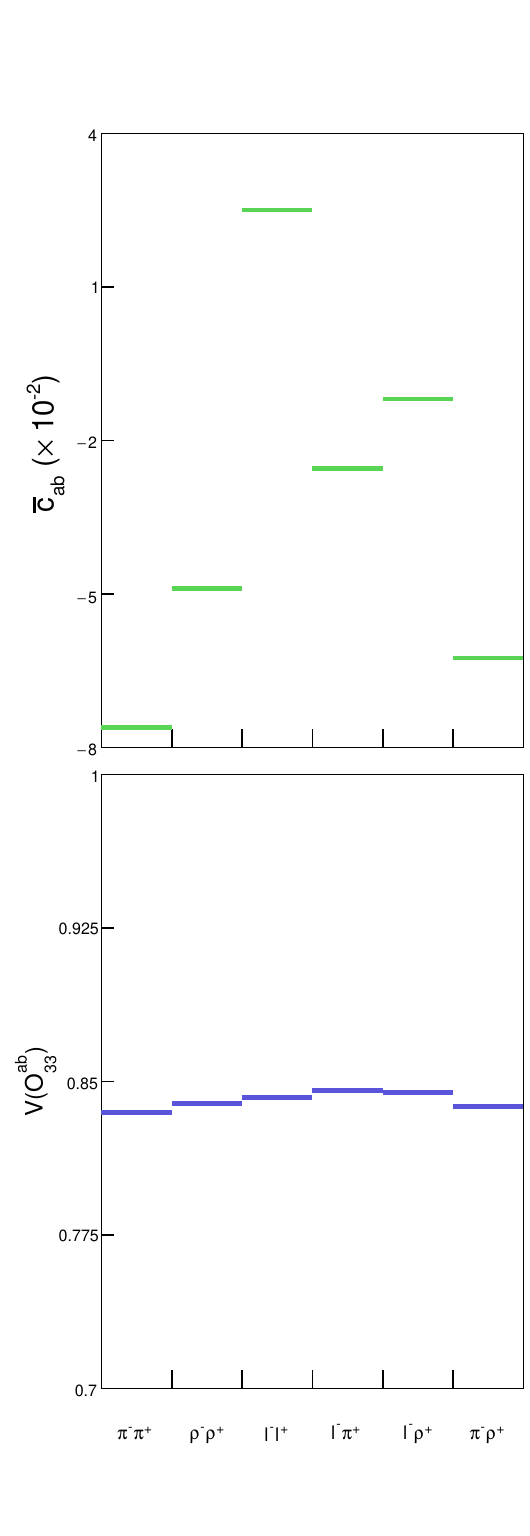}
\caption{Simple observables $T^{ij}$(left) and $\hat O^{ij}$(right) at $\sqrt s=3.686$ GeV. The upper panel shows our results for $c_{ab}$ and $\bar c_{ab}$ defined in eq.(\ref{eq:cab}) while the lower panel exhibits the variances of the observables defined in eq.(\ref{eq:covariance}). The x-axis represents the various decay modes. For the modes where leptons are involved the cross section is the sum of the respective cross sections for $\ell = e$ and $\mu$.  }
\label{fig:senario1simple}
\end{figure}

Since we consider the normalized expectation values of observables, the resonance effects where the propagator receives an enhancement  at the resonances $\psi(2S)$,  $\Upsilon(4S)$ and the Z boson resonance, enters only through the number of $\tau$ lepton events, which serves as  an input parameter.
Fig.\ref{fig:senario1simple} shows the results for the expectation values of the simple observables $T^{ij}$ at the left panel and $\hat O^{ij}$ at the right panel in the Scenario I assuming $ 3.5\times 10^6$ produced $\tau$ pairs at $\sqrt s=3.686$ GeV. The upper panel is the results of the coefficient matrix defined in eq.(\ref{eq:cab}), the values of which are expected to be large; while the lower panel is the variances of the observables in eq.(\ref{eq:covariance}) and we expect their values to be small. The x-axis represents various decay modes of $\tau^+\tau^-$.
 The upper and lower panel can be seen as signal and background strength. As seen from the values in the upper panel , the $\pi^+\pi^-$ channel has the highest sensitivities to both $\text{Re}[d_{\tau}^\gamma]$ and $\text{Im}[d_{\tau}^\gamma]$. This is due to the high spin analyzing power of the $\pi$. Furthermore, channels with pions in the final state are more sensitive to $\text{Re}[d_{\tau}^\gamma]$ than leptonic channels, while the $\hat O^{33}$ in $\rho^+\rho^-$ channel is more relevant to $\text{Im}[d_{\tau}^\gamma]$ than leptonic and semi-leptonic decay channel of $\tau^+\tau^-$. 
From the lower panel, the measurement errors for $\text{Re}[d_{\tau}^\gamma]$ in  $\pi^+\pi^-$ and $\pi^-\rho^+$ are also large. The errors differences for measuring $\text{Im}[d_{\tau}^\gamma]$ across various decay channels are relatively small.
 
\begin{figure}[!h]
\centering
\includegraphics[scale=0.45]{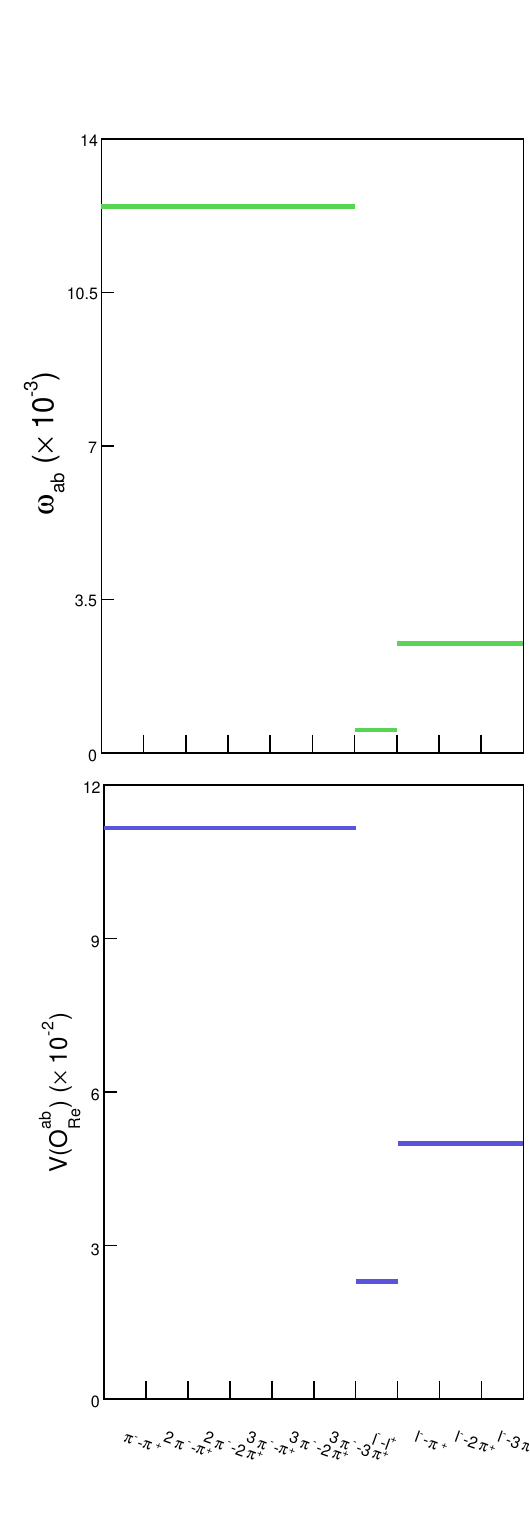}
\hspace{0.4in}
\includegraphics[scale=0.45]{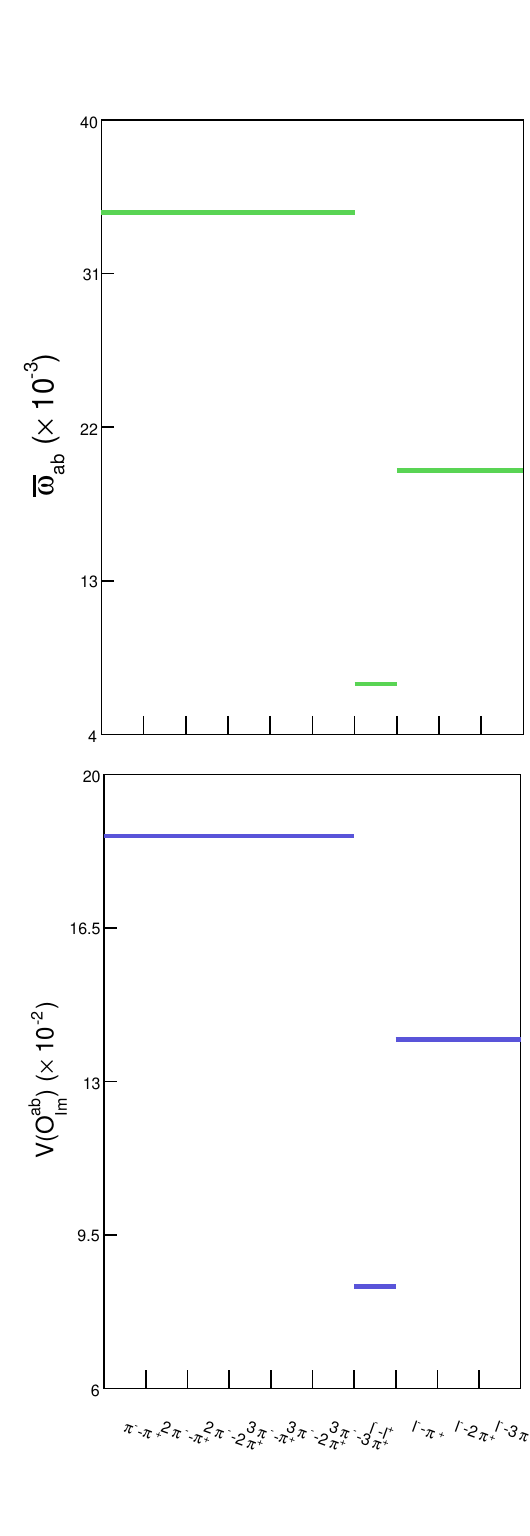}
\caption{Optimal observables $\oo_{Red_{\tau}}$(left) and $\oo_{Imd_{\tau}}$(right) at  $\sqrt s=3.686$ GeV. The upper and lower panels show quantities analogous to Fig.\ref{fig:senario1simple}.}
\label{fig:senario1optimal}
\end{figure}

\begin{figure}[!h]
\centering
\includegraphics[scale=0.4]{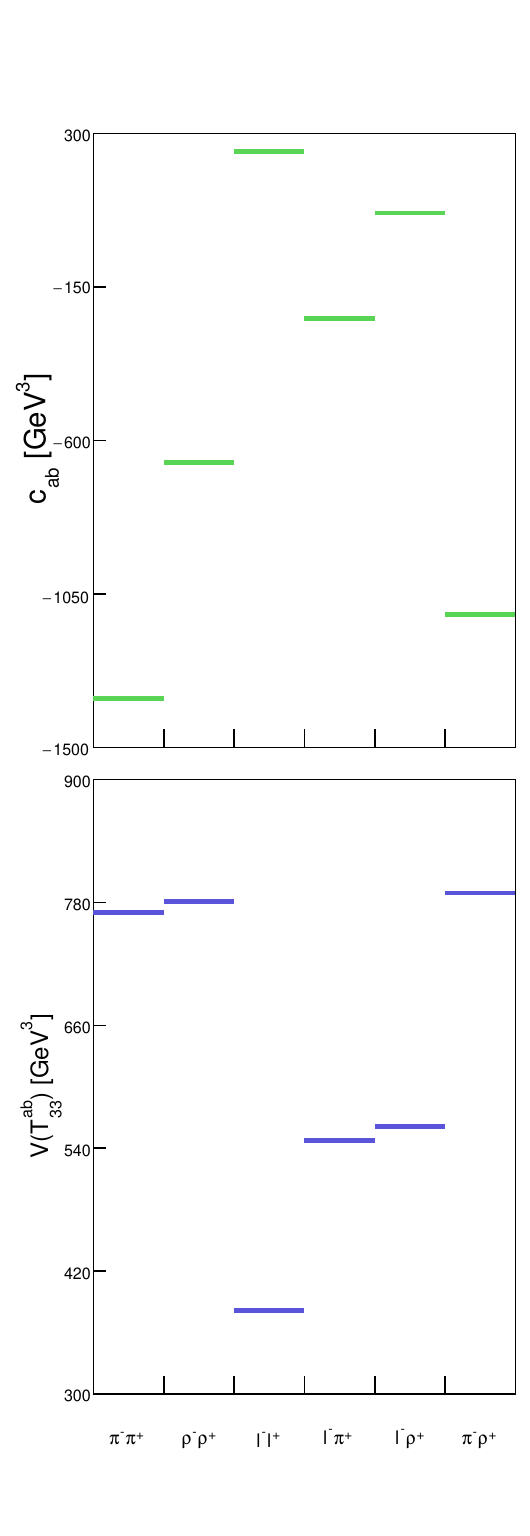}
\hspace{0.4in}
\includegraphics[scale=0.4]{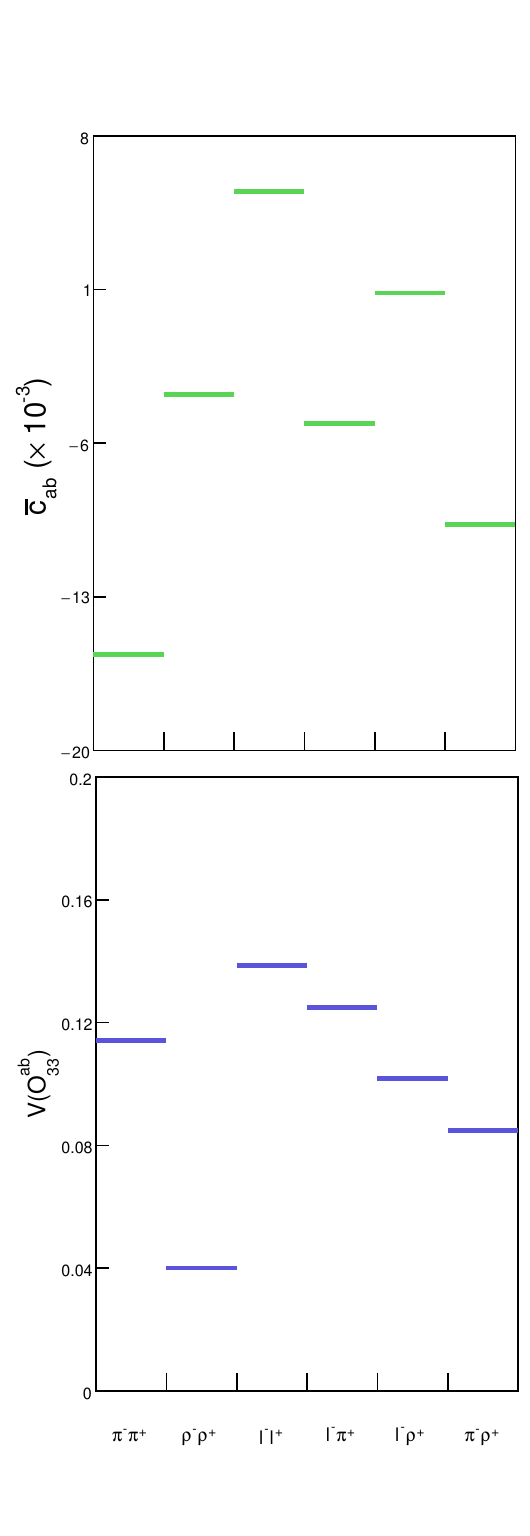}
\caption{ The same as the Fig.\ref{fig:senario1simple}, but here the c.m. energy is $\sqrt s = m_Z$.}
\label{fig:TO33}
\end{figure}

\begin{figure}[!h]
\centering
\includegraphics[scale=0.4]{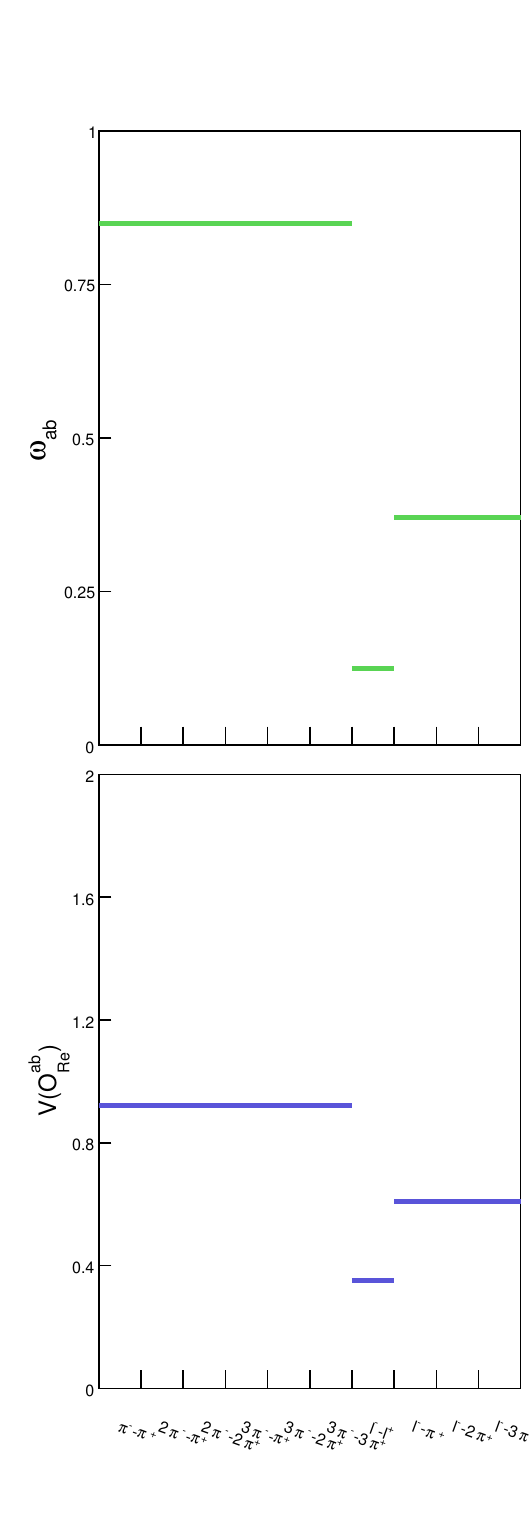}
\hspace{0.4in}
\includegraphics[scale=0.4]{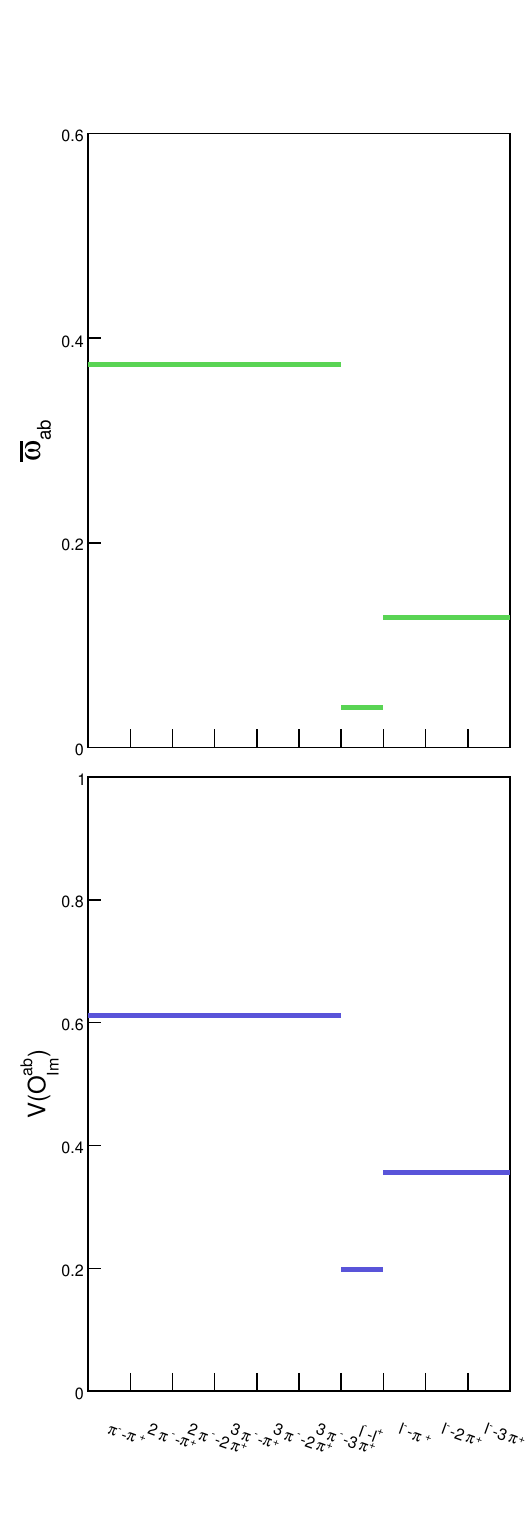}
\caption{The same as the Fig.\ref{fig:senario1optimal}, but here the c.m. energy is $\sqrt s = m_Z$.}
\label{fig:Oreimz}
\end{figure}

 Fig.\ref{fig:senario1optimal} presents the results for the expectation values of the optimal observables $\oo_{Red_{\tau}}$ and $\oo_{Imd_{\tau}}$. The upper panel shows that the sensitivity of these two observables in double-leptonic decay of $\tau$ pair are much lower than other decay modes. As shown in the lower panels, these optimal observables exhibit smaller statistical uncertainties than the simple observables in Fig.\ref{fig:senario1simple}. Analogous behaviors are also observed for the Scenario II and III. For the WDM of the $\tau$ lepton (Scenario IV), Fig.\ref{fig:TO33} displays the results for simple observables. The $\rho^+\rho^-$ channel shows sensitivity to $\oo_{Red_{\tau}}$, complementing the preciously noted $\pi^+\pi^-$ and $\pi^-\rho^+$ channels. The absolute values of coefficient $c_{ab}$ are markedly larger than those in $\gamma$ coupling. The statistical uncertainties of the optimal observables in Fig.\ref{fig:Oreimz} are smallest at $\sqrt s = m_Z$.

Combining the statistical errors of the expectation values of the observables with the signal strength, we compute the 1 s.d. sensitivities shown in Table \ref{table:sd1} for various scenarios. The table demonstrates significant improvements in precision when using the optimal observables compared to the simple observables. At $\sqrt s = 3.686$ GeV, the sensitivity reaches  $4.87\times 10^{-17}$ $ecm$ for the real component  of EDM and $2.40\times 10^{-17}$ $ecm$ for the imaginary component of EDM. When the energy increases to $\sqrt s = 10.58$ GeV,  the  $\delta \text{Re}d_{\tau}^{\gamma}$ and $\delta \text{Im}d_{\tau}^{\gamma}$ improve by approximately three orders of magnitude,  reaching $10^{-20}$ $ecm$. Our results are consistent with ref.\cite{Bernreuther:2021uqm}, with the slight discrepancy attributed to differences in the input parameters. 
 For the WDM of $\tau$ the lepton, the numbers show that the sensitivity to $d^Z_{\tau}$ at CEPC can reach the level of $10^{-21}$ $ecm$ using $\oo_{Red_{\tau}}$ and $\oo_{Imd_{\tau}}$. This is sufficient to be sensitive to certain BSM that predict values for $d_{\tau}^Z$ of the order $10^{-19}$ $ecm$ (See, e.g. \cite{Huang:1996jr}).  Compared to the simple observables, the sensitivity to $\text{Re}[d^Z_{\tau}]$ is improved by more than a factor three with $\oo_{Red_{\tau}}$ while the sensitivity to $\text{Im}[d^Z_{\tau}]$ is improved by more than a factor twenty.
 
\begin{table}[h]
\renewcommand{\arraystretch}{1.5}
\caption{1 s.d. sensitivity on $\red_{\tau}^V$ and $\imd_{\tau}^V$ in the combined decay channels.}\label{table:sd1}
\begin{center}
\setlength{\tabcolsep}{5mm}{
\begin{tabular}{l l c l c }
\hline \hline 
	&\multicolumn{2}{l}{$ \delta \red_{\tau}^{V}$ [$ecm$]} 	&	\multicolumn{2}{l}{$ \delta \imd_{\tau}^{V}$ [$ecm$]} \\ \hline
	&				$E_{ab}[T_{33}]$	&$E_{ab}[\oo_{Red_{\tau}}]$	&$E_{ab}[\hat{O}_{33}]$	&$E_{ab}[\oo_{Imd_{\tau}}]$	\\
$\sqrt s$=3.686 GeV	&	\multirow{2}{*}{$4.11\times 10^{-16}$}	&\multirow{2}{*}{$4.87\times 10^{-17}$} 	& \multirow{2}{*}{$2.97\times 10^{-16}$}	&\multirow{2}{*}{$2.40\times 10^{-17}$}	\\
$(N_{\tau^+\tau^-} =	3.5 \times10^6)$	&	\\				
\hline 
$\sqrt s$=5.6 GeV	   &\multirow{2}{*}{$1.78\times 10^{-17}$}	&\multirow{2}{*}{$2.93\times 10^{-18}$}	& \multirow{2}{*}{$6.52\times 10^{-18}$}	&\multirow{2}{*}{$1.42\times 10^{-18}$}	\\	
$(N_{\tau^+\tau^-} =	5.5 \times10^7)$	&	\\
\hline
$\sqrt s$=10.58 GeV	   &\multirow{2}{*}{$ 2.93\times 10^{-19}$}	&\multirow{2}{*}{$ 9.42\times 10^{-20}$}	& \multirow{2}{*}{$ 5.08\times 10^{-20}$}	&\multirow{2}{*}{$ 2.43\times 10^{-20}$}	\\	
$(N_{\tau^+\tau^-} =	5.5 \times10^{10})$	&	\\
\hline
$\sqrt s$= Z-pole energy	   &\multirow{2}{*}{$ 3.67\times 10^{-21}$}	&\multirow{2}{*}{$1.12\times 10^{-21}$}	& \multirow{2}{*}{$4.36\times 10^{-20}$}	&\multirow{2}{*}{$1.78\times 10^{-21}$}	\\	
$(N_{\tau^+\tau^-} =	1.2 \times10^{11})$	&	\\
\hline \hline 
\end{tabular}
}
\end{center}
\end{table}

\subsection{AMDM Results}
Considering the high statistics of the $\tau$ pairs, we study the AMDM of $\tau$ lepton at $\sqrt s = 10.58$ GeV in the Belle-II experiment. We assume that $N_{\tau\tau} = 4.5\times 10^{10}$ $\tau$-pair events can be eventually recorded. After diagonalizating the coefficient matrix , we calculate the expectation values of the simple observables $T_a^{ij}$, $\hat{O}^{ij}_a$ and optimal observables $\oo_{Rea_{\tau}}$, $\oo_{Ima_{\tau}}$. 
 In Table \ref{table:g-2simple} and Table \ref{table:g-2}, we give results of the 1 s.d. sensitivities to the real and imaginary components of the $\tau$ AMDM calculated using the differential cross section at SM tree level. The optimal observables in this table are defined by tree level $S_{0}$. Comparing the data from those two tables, the sensitivity of optimal observables is of the order of $10^{-5}$, an order of magnitude less than that of the simple observable. 
 Thanks to a large branching ratio and spinor analyzing power, the precision resulting from $\tau^- \to \nu\pi^0\pi^-$, $\tau^+ \to \nu\pi^0\pi^+$ and $\tau^- \to \nu\pi^0\pi^-$, $\tau^+ \to \bar{\nu}\nu\ell^+$ final state is very high and can reach $4.23 \times 10^{-5}$ and $3.71 \times 10^{-5}$, respectively. 
 Combining these 1 s.d. listed in Table \ref{table:g-2simple} and Table \ref{table:g-2} in quadratures,  one gets the 1 s.d. sensitivity  $\delta \rea^{\gamma}_{\tau}$ and  $\delta \ima^{\gamma}_{\tau}$ shown in Table \ref{table:asd1}. The sensitivity of optimal observables yields a factor of three improvement in precision compared to simple one for the real component of $a^{NP}_{\tau}$, while for the imaginary component of $a^{NP}_{\tau}$, there is an improvement by a factor of about seven. The sensitivity of $\oo_{Ima_{\tau}}$ reaches $1.76\times 10^{-5}$ and of $\oo_{Rea_{\tau}}$ reaches $1.88\times 10^{-5}$.
 
 \begin{table}[h]
\caption{Simple observables $T_a^{33}$, $\hat{O}_a^{33}$ for the  $e^+e^- \to \gamma^* \to \tau\tau$ at Belle-II with $\sqrt s = 10.58$ GeV. For the decay modes where leptons are involved the cross section is the sum of the respective cross sections for $\ell = e$ and $\mu$.  }\label{table:g-2simple}
\begin{center}
\begin{tabular}{c c c c c c c c}
\hline \hline
 & 	& $\kappa_{\gamma}$ & $\sqrt{V^{'}(T_a^{33})}$  	& $\delta \ima^{\gamma}_{\tau}$	 & $\bar{\kappa}_{\gamma}$	& $\sqrt{V^{'}(\hat{O}_a^{33})}$	&$\delta \rea^{\gamma}_{\tau}$ \\ 
$\tau^- \to$ 		& $\tau^+ \to$	& 			 &		&($10^{-4}$)	 &		&		&($10^{-4} $)	\\
\hline
$\nu \pi^-$			& $\bar{\nu} \pi^+$	 &1.21    &3.34    &3.61    &-0.200    &0.640    &4.19\\                                      
\hline
\\
$\nu \rho^-$		&$\bar{\nu} \rho^+$	&0.295    &2.61     &4.91    &-0.301    &0.612    &1.13\\
\hline
\\
$\nu \bar{\nu}\ell^-$	&$\bar{\nu} \nu \ell^+$  &-0.099		&1.69     &6.96     &-0.222     &0.651        &1.19\\
\hline
\\
$\nu \bar{\nu}\ell^-$	&$\bar{\nu} \pi^+$	&-0.236     &2.77    &6.05    &-0.238    &0.646    &1.40\\
\hline
\\
$\nu\bar{\nu} \ell^-$		&$\bar{\nu} \rho^+$	 &-0.403     &2.57     &2.14     &-0.264    &0.637     &0.811 \\
\hline
\\
$\nu \pi^-$		&$\bar{\nu} \rho^+$	 &-0.903     &3.18     &2.12     &0.257    &0.630     &1.48 \\
\hline
\end{tabular}
\end{center}
\end{table}

\begin{table}[h]
\caption{The same as  Table.\ref{table:g-2simple}, but for optimal observables $\oo_{Rea_{\tau}}$ and $\oo_{Ima_{\tau}}$. }\label{table:g-2}
\begin{center}
\begin{tabular}{c c c c c c c c}
\hline \hline
 & 	& $\xi_{\gamma}$ & $\sqrt{V^{'}(\oo_{Rea_{\tau}})}$  	& $\delta \rea^{\gamma}_{\tau}$	 & $\xi^{\prime}_{\gamma}$	& $\sqrt{V^{'}(\oo_{Ima_{\tau}})}$	&$\delta \ima^{\gamma}_{\tau}$ \\ 
$\tau^- \to$ 		& $\tau^+ \to$	& 			 &		&($10^{-5}$)	 &		&		&($10^{-5} $)	\\
\hline
$\nu \pi^-$		& $\bar{\nu} \pi^+$	 &0.191    &0.437    &9.97    &0.186    &0.432    &10.1\\                                      
\hline
\\
$\nu \pi^0\pi^-$		&$\nu \pi^0\pi^+$	&0.191    &0.437     &4.22    &0.186    &0.432    &4.23\\
\hline
\\
$\nu \pi^-\pi^-\pi^+$	&$\nu \pi^+\pi^+\pi^-$  &0.191	&0.437     &11.5     &0.186     &0.432        &11.7 \\
\hline
\\
$\nu \pi^0\pi^-$		&$\bar{\nu} \pi^+$	&0.191     &0.437    &4.59    &0.186    &0.432    &4.65\\
\hline
\\
$\nu \pi^-\pi^-\pi^+$	&$\bar{\nu} \pi^+$	 &0.191     &0.437     &7.60     &0.186    &0.432     &7.70 \\
\hline
\\
$\nu \pi^-\pi^-\pi^+$	&$\nu \pi^0\pi^+$	 &0.191     &0.437     &4.94     &0.186    &0.432    &5.01    \\
\hline
\\
$\nu \bar{\nu} \ell^-$	&$\bar{\nu} \nu \ell^+$	&0.036   &0.190    &7.12    &0.031    &0.177    &7.65\\
\hline
\\
$\nu \pi^-$			&$\bar{\nu} \nu \ell^+$	 &0.060    &0.244    &7.04    &0.091   &0.301    &5.70\\
\hline
\\
$\nu \pi^0\pi^-$		&$\bar{\nu} \nu \ell^+$	 &0.060    &0.244    &4.58    &0.091    &0.301   &3.71\\
\hline
\\
$\nu \pi^-\pi^-\pi^+$	&$\bar{\nu} \nu \ell^+$   &0.060    &0.244    &7.58    &0.091    &0.301    &6.14\\
\hline
\end{tabular}
\end{center}
\end{table}

 \begin{table}[h]
\renewcommand{\arraystretch}{1.5}
\caption{1 s.d. sensitivity on $\rea_{\tau}$ and $\ima_{\tau}$ obtained from the combined decay channels.}\label{table:asd1}
\begin{center}
\setlength{\tabcolsep}{5mm}{
\begin{tabular}{ l c l c }
\hline \hline 
\multicolumn{2}{l}{$ \delta \rea_{\tau}^{\gamma}$ } 	&	\multicolumn{2}{l}{$ \delta \ima_{\tau}^{\gamma}$ } \\ \hline
				$E_{ab}[\hat{O}_a^{33}]$		&$E_{ab}[\oo_{Rea_{\tau}}]$	&$E_{ab}[T_a^{33}]$		&$E_{ab}[\oo_{Ima_{\tau}}]$	\\
				$4.98\times 10^{-5}$		&$1.88\times 10^{-5}$ 		&$1.28\times 10^{-4}$		&$1.76\times 10^{-5}$	\\			
\hline \hline 
\end{tabular}
}
\end{center}
\end{table}

 As is well-known, the expectation values of $\oo_{\rea_{\tau}}$ and $\oo_{\ima_{\tau}}$ receive contributions from both SM interactions and potential NP. If the effect of $a_{\tau}^{SM}$ on the expectation value exceeds the statistical uncertainty of the measurement, then the SM contribution can be constrained by measuring $E^{ab}[\oo]$.  $E^{ab}[\oo]$ can be fitted with a linear function of $a_{\tau}^{NP}$
\begin{align}
E^{ab}[\oo_i] = \xi^i a^{NP}_{\tau,i} + b^i, ~(i=Rea_{\tau}, Ima_{\tau}) ,
\end{align} 
where $\xi^i$ is the slope  and $b^i$ corresponds to the SM expectation value of $\oo_i$. For the decay channel $\tau^- \to \nu\pi^0\pi^-, \tau^+ \to \nu\pi^0\pi^+$, tree level differential cross section gives $E_0[\oo_{Rea_{\tau}}]=0.9539$ with a slope of 0.1915. In the SM, 
the differential cross section related to $a_{\tau}^{SM}$ can be factorized into 
$a_{\tau}^{SM} S_{1,Rea_{\tau}}$
which has the same effective vertex as $a_{\tau}^{NP}$. Using the SM prediction value of $a_{\tau}$ $a_{\tau}^{SM} = (1.177171 \pm 0.000039) \times 10^{-3}$, we derive the bounds on $\xi^{\rea}$,  $b^{\rea}$ and sensitivity by varying $a_{\tau}^{SM}$ within its uncertainty range, shown in Table \ref{table:atau}. Our results indicate that the observables defined as $S_{1,Rea_{\tau}}/S_{0}$ can not identify $a_{\tau}^{NP}$ from $a_{\tau}^{SM}$ within the current sensitivity limits. 

\begin{table}[h]
\renewcommand{\arraystretch}{1.5}
\caption{The sensitivity to $a_{\tau}^{SM}$ by measuring $\oo_{Rea_{\tau}}$.}\label{table:atau}
\begin{center}
\setlength{\tabcolsep}{5mm}{
\begin{tabular}{ c c c c }
\hline \hline 
$a_{\tau}^{SM}$($\times 10^{-3}$)		&$E_0[\oo_{Rea_{\tau}}]$ 		&$\xi^{\rea_{\tau}}$ 		&$\delta\rea_{\tau}$ \\ \hline
0		&0.9539						&0.1915				&$4.225 \times 10^{-5}$	\\
1.17721	&0.9542						&0.1911				&$4.237 \times 10^{-5}$		\\
1.177171	&0.9542						&0.1911				&$4.237\times 10^{-5}	$\\
1.17713	&0.9542						&0.1911				&$4.237\times 10^{-5}$\\
\hline \hline 
\end{tabular}
}
\end{center}
\end{table}

\section{Summary}\label{sec5}
The precise measurement of the $\tau$ lepton's electric(weak) dipole moment and anomalous magnetic dipole moment  provide a powerful probe for testing the SM and uncovering NP. Electron-positron colliders offer an ideal place for these studies.  With ongoing advancements in both the collider energy and luminosity, $e^+e^-$ facilities will produce large datasets of $\tau^{+}\tau^{-}$, hence the search for $\tau$ E(W)DM and AMDM will enter a precision era.

In this work, we study $\tau$ pair production in $e^+e^-$ collisions with subsequent decays of $\tau^{\pm}$ to (semi)leptonic or hadronic channels in various scenarios. For all one-prong and multi-prongs decay channels considered, we take the full kinematic information of the final-states system into account by incorporating the respective differential $\tau$ lepton decay density matrix into its production density matrix. The one standard deviation sensitivity to the real and imaginary components of E(W)DM and AMDM are estimated through both simple and optimal observables, with comprehensive analysis of their expectation values and covariance.

At BEPC II, optimal observables yield an order of magnitude improvement in precision over simple observables for measurements of real and imaginary components of the $\tau$ EDM at center-of-mass energies of 3.686 GeV and 5.6 GeV. Specifically, with $3.5\times 10^6$ $\tau$ pair production at $\sqrt s = 3.686$ GeV, the sensitivities derived from optimal observables read $\delta \red_{\tau}= 4.87\times 10^{-17}$ $ecm$ and $\delta \imd_{\tau}= 2.40\times 10^{-17}$ $ecm$, with further enhancement at higher energies. 
For Belle-II, our results project the precisions of  $9.42\times10^{-20}$ $ecm$ and $2.43\times10^{-20}$ $ecm$ for $\red_{\tau}$ and  $\imd_{\tau}$ respectively, while for AMDM  $\delta \rea_{\tau}=1.88\times 10^{-5}$ and $\delta \ima_{\tau}= 1.76\times 10^{-5}$.  CEPC has the capability to measure $\tau$ lepton WDM with precisions reach the level of $10^{-21}$ $ecm$, with specific values of $\delta \red_{\tau}=1.12\times 10^{-21} and~  \delta \imd_{\tau}= 1.78\times 10^{-21}$ at the Z mass energy, which is sufficient to be sensitive to certain BSM models. Precision measurement of the $\tau$ lepton's electromagnetic moments using optimal observables at current and future electron-positron colliders will provide significant information about physics beyond the standard model, offering high sensitivity to leptonic CP violation.

\section*{Acknowledgements}
The authors sincerely thank Professor Werner Bernreuther, Long Chen for their invaluable discussions and comments of this manuscript,
and the members of the Institute of theoretical physics of Shandong University for helpful communications.
This work is supported in part by National Natural Science Foundation of China 
under the Grants  No. 12235008, No. 12321005 and No. 12405121.

\section*{Appendix}

In this Appendix, we give the production spin density matrix for $e^+e^- \to \gamma^* /Z \to \tau^+\tau^-$ process defined in eq.(\ref{eq:densityM}). $V_{\ell}^{\gamma}$, $V_{\ell}^{Z}$, $A_{\ell}^{\gamma}$, $A_{\ell}^{Z}$, where $\ell$ represents $e$ and $\tau$ lepton respectively, are the vector and axial-vector couplings of the electron and the $\tau$ lepton to the photon and Z boson:
\begin{align}
&V^{\gamma}_{\ell} = Q_{\ell} e, ~~ V^{Z}_{\ell} =  \frac{e (T_{3\ell} - 2 Q_{\ell} \sin^2 \theta_w)}{2\sin \theta_w \cos\theta_w}, \nonumber \\
 &A^{\gamma}_{\ell} = 0,
~~A^{Z}_{\ell} = \frac{e T_{3\ell} }{2\sin \theta_w \cos\theta_w}.
\end{align}
The SM contribution that results from photon exchange can be written

\begin{eqnarray}\label{Eq.chip0}
\chi_{0,\text{tree}}^\gamma &= & \frac{s (V_{e}^{\gamma})^2 (V_{\tau}^{ \gamma})^2}{Dq}  \left\{ [k_0^2 + m_\tau^2 +|\kk|^2 (\hk \cdot\hp)^2] -({\sip}\cdot{\ssim})|\kk|^2[1-(\hk \cdot \hp)^2] \right. \nonumber \\
& & \left. + 2 (\hk \cdot \sip)(\hk \cdot \ssim)[|\kk|^2 +(k_0-m_\tau)^2 (\hk \cdot \hp)^2]  + 2 k_0^2 (\hp \cdot \sip)(\hp \cdot \ssim) \right. \nonumber\\
& & \left.  - 2 k_0(k_0-m_\tau) (\hk \cdot \hp)[ (\hk \cdot \sip)(\hp \cdot \ssim)
 + (\hk \cdot \ssim)(\hp \cdot \sip)] \right \} \, ,  
\end{eqnarray}
where $\hat{\bold{p}}$ and $\hat{\bold{k}}$  denote the unit vector of the electron and $\tau$ lepton in \ee ~ c.m. frame and $|\bold{k}|$ is the magnitude of the $\tau$ lepton's three-momenta in the same frame. We use the following notation
\begin{eqnarray}
\sip = \sigma_{\alpha \alpha'}\delta_{\beta \beta'}, ~~~ &\ssim=\delta_{\alpha \alpha'}\sigma_{\beta \beta'}.\\
Dq = \frac{1}{q^4}, ~~~&Dzq = \frac{1}{(q^2-m_Z^2)^2 - \Gamma_Z^2 m_Z^2}
\end{eqnarray}
with $q=\sqrt s$. The interference terms between SM matrix elements and EDM relevant matrix elements for the photon propagator reads
\begin{eqnarray}\label{Eq.chipred}
\chi_{Red}^\gamma & = &  2  \frac{s^{3/2} (V_e^\gamma)^2 (V_\tau^\gamma)  |\kk|}{Dq} \left\{ -[m_\tau + (k_0-m_\tau) (\hk \cdot\hp)^2](\sip \times \ssim)\cdot\hk \right. \nonumber \\
 & & \left. + k_0(\hk \cdot\hp) (\sip \times \ssim)\cdot\hp \right\} \, , 
\end{eqnarray}

\begin{eqnarray}\label{Eq.chipimd}
\chi_{Imd}^\gamma & = & - 2 \frac{s^{3/2}  (V_e^\gamma)^2 (V_\tau^\gamma) |\kk|}{Dq} \left\{ -[m_\tau + (k_0-m_\tau) (\hk \cdot\hp)^2](\sip - \ssim)\cdot\hk \right. \nonumber \\
 & & \left. + k_0(\hk \cdot\hp) (\sip - \ssim)\cdot\hp \right\} \, .
\end{eqnarray}
Results for the SM contribution and the interference contribution for the Z propagator are shown in the following
\begin{eqnarray}\label{Eq.chiz0}
\chi^Z_{0,\text{tree}} &= & \frac{s ((V_{e}^{Z})^2 + (A_{e}^{Z})^2)}{Dzq} \bigg\{
(V_{\tau}^Z)^2  \left\{ [k_0^2 + m_\tau^2 +|\kk|^2 (\hk \cdot\hp)^2] -({\sip}\cdot{\ssim})|\kk|^2[1-(\hk \cdot \hp)^2] \right. \nonumber \\
& & \left. + 2 (\hk \cdot \sip)(\hk \cdot \ssim)[|\kk|^2 +(k_0-m_\tau)^2 (\hk \cdot \hp)^2]  + 2 k_0^2 (\hp \cdot \sip)(\hp \cdot \ssim) \right. \nonumber\\
& & \left.  - 2 k_0(k_0-m_\tau) (\hk \cdot \hp)[ (\hk \cdot \sip)(\hp \cdot \ssim)
 + (\hk \cdot \ssim)(\hp \cdot \sip)] \right \} \nonumber\\
&& + (A_{\tau}^Z)^2 |\kk|^2
 \left\{ 1+(\hk \cdot \hp)^2 + (\sip \cdot \ssim)(1-(\hk \cdot \hp)^2)-2(\hp \cdot \sip)(\hp \cdot \ssim) \right. \nonumber\\
& &\left. +2\hk\cdot\hp((\hk \cdot \sip)(\hp \cdot \ssim)+(\hk \cdot \ssim)(\hp \cdot \sip))\right\} \nonumber\\
& & +2 V_{\tau}^Z A_{\tau}^Z |\kk| \left\{\hk \cdot (\sip +\ssim)(k_0+(k_0-m)(\hk \cdot \hp)^2)+m(\hk \cdot \hp)\hp \cdot (\sip +\ssim)\right\} 
\bigg\}\nonumber\\
&& + \frac{2 V_{e}^Z A_{e}^Z}{Dzq} \bigg\{ (V_{\tau}^Z)^2 2 k_0 \left\{m \hp \cdot(\sip +\ssim)+(k_0-m)(\hk \cdot \hp)\hk \cdot(\sip +\ssim)
\right\}\nonumber\\
&&+ (A_{\tau}^Z)^2 2 |\kk|^2 (\hk \cdot \hp) \hk \cdot (\sip +\ssim) + V_{\tau}^Z A_{\tau}^Z |\kk| \left\{m((\hk \cdot \ssim)(\hp \cdot \sip)+(\hk \cdot \sip)(\hp \cdot \ssim)) \right.  \nonumber\\
&&\left. + 2 k_0 (\hk \cdot \hp) + 2(k_0-m)(\hk \cdot \hp)(\hk \cdot \sip)(\hk \cdot \ssim)
\right\}
\bigg\}
 \, ,  
\end{eqnarray}

\begin{eqnarray}\label{Eq.chizred}
\chi_{Red}^Z & = & \frac{2 s^{3/2} ((V_{e}^{Z})^2 + (A_{e}^{Z})^2)) |\kk|}{Dzq} \bigg\{
V_{\tau}^Z \left\{ -[m_\tau + (k_0-m_\tau) (\hk \cdot\hp)^2](\sip \times \ssim)\cdot\hk \right. \nonumber \\
 & & \left. + k_0(\hk \cdot\hp) (\sip \times \ssim)\cdot\hp \right\} +A_{\tau}^Z |\hk| \left\{ -(\hk \cdot\hp)(\hk \times (\sip -\ssim))\cdot \hp   \right\}  \bigg\}\nonumber \\
  & &+\frac{4 s^{3/2} V_{e}^Z A_{e}^Z |\kk|}{Dzq}\left\{A_{\tau}^Z ((\hk \cdot \sip)(\hk \times \ssim)\cdot \hp-(\hk \cdot \ssim)(\hk \times \sip)\cdot \hp)  \right. \nonumber \\
&& \left.  - V_{\tau}^Z(k_0(\hk \times (\sip -\ssim))\cdot \hp)
 \right\},
\end{eqnarray}

\begin{eqnarray}\label{Eq.chizimd}
\chi_{Imd}^Z & = &   \frac{2 s^{3/2} ((V_{e}^{Z})^2 + (A_{e}^{Z})^2)) |\kk|}{Dzq} \bigg\{ -V_{\tau}^Z \left\{ -[m_\tau + (k_0-m_\tau) (\hk \cdot\hp)^2](\sip - \ssim)\cdot\hk \right. \nonumber \\
 & & \left. + k_0(\hk \cdot\hp) (\sip - \ssim)\cdot\hp \right\} \nonumber \\
 & & -A_{\tau}^Z |\hk| \left\{  -(\hk \cdot\hp)((\hk \cdot \sip)(\hp \cdot \ssim)-(\hk \cdot \ssim)(\hp \cdot \sip))\right\}
\bigg\}\nonumber \\
& &+\frac{4 s^{3/2} V_{e}^Z A_{e}^Z |\kk|}{Dzq} \left\{V_{\tau}^Z k_0((\hk \cdot \sip)(\hp \cdot \ssim)-(\hk \cdot \ssim)(\hp \cdot \sip))\right. \nonumber \\
&& \left.-A_{\tau}^Z (\hp\cdot(\sip -\ssim) - (\hk \cdot\hp) \hk (\sip -\ssim))\right\}
 \, .
\end{eqnarray}
The above SM and interference matrix elements are the same as the results in ref.\cite{Bernreuther:1993nd}.

The interference terms between SM matrix elements and AMDM matrix elements for the photon 
is given by $\chi_{Ima}^\gamma$ and $\chi_{Rea}^\gamma$.
\begin{eqnarray}\label{Eq.chipima}
\chi_{Ima}^\gamma & = & - 2 \frac{s  (V_e^\gamma)^2 (V_\tau^\gamma) e |\kk|^2}{Dq} (\hk \cdot\hp) ((\hp \times \sip) \cdot  \hk  +  (\hp \times \ssim) \cdot  \hk)
\, , 
\end{eqnarray}

\begin{eqnarray}\label{Eq.chiprea}
\chi_{Rea}^\gamma & = &  2  \frac{s (V_e^\gamma)^2 (V_\tau^\gamma) e }{Dq} \left\{ -(\hk \cdot \hp)(k_0-m)^2((\hk \cdot \ssim)(\hp \cdot \sip) +(\hk \cdot \sip)(\hp \cdot \ssim))   \right. \nonumber \\
&& \left. + 2(\hk \cdot \hp)(k_0-m)^2(\hk \cdot \sip)(\hk \cdot \ssim)-2k_0m(1+(\hp \cdot \sip)(\hp \cdot \ssim))
 \right\} \, .
\end{eqnarray}

\end{document}